\documentclass[prb,reprint,twocolumn,showpacs,superscriptaddress]{revtex4-1}
\usepackage{amsmath,amssymb,bm,mathrsfs,graphicx, braket, times}
\usepackage[colorlinks=true,citecolor=blue,linkcolor=blue]{hyperref}
\usepackage{lipsum}
\usepackage{longtable}
\usepackage{multirow}
\usepackage[all,cmtip]{xy}
\usepackage[normalem]{ulem}
\usepackage[usenames,dvipsnames]{color}			

\renewcommand{\vec}[1]{\bm{#1}}

\newcommand{\CGP}{Ca$_2$Pt$_2$Ga}

\begin{document}

\title{
Topological Materials Discovery using Electron Filling Constraints
}

\author{Ru Chen}\thanks{These two authors contributed equally to this work.}
\affiliation{Department of Physics, University of California, Berkeley, CA 94720, USA}
\affiliation{Molecular Foundry, Lawrence Berkeley National Laboratory, Berkeley, California 94720, USA}

\author{Hoi Chun Po}\thanks{These two authors contributed equally to this work.}
\affiliation{Department of Physics, University of California, Berkeley, CA 94720, USA}
\affiliation{Department of Physics, Harvard University, Cambridge MA 02138, USA}

\author{Jeffrey B. Neaton}
\affiliation{Department of Physics, University of California, Berkeley, CA 94720, USA}
\affiliation{Molecular Foundry, Lawrence Berkeley National Laboratory, Berkeley, California 94720, USA}
\affiliation{Kavli Energy Nanosciences Institute, Berkeley, California 94720, USA}

\author{Ashvin Vishwanath}
\affiliation{Department of Physics, University of California, Berkeley, CA 94720, USA}
\affiliation{Department of Physics, Harvard University, Cambridge MA 02138, USA}

\begin{abstract}
Nodal semimetals, materials systems with nodal-point or -line Fermi surfaces, are much sought after for their novel  transport and topological properties. Identification of experimental materials candidates, however, has mainly relied on exhaustive database searches.  Here we show how recent studies on the interplay between electron filling and nonsymmorphic space-group symmetries can guide the search for nodal semimetals which are `filling-enforced'. We recast the previously-derived constraints on the allowed electron fillings for band insulators in the 230 space groups into a new form, which enables effective screening of materials candidates based solely on their (1) space group, (2) electron count in the formula unit, and (3) multiplicity of the formula unit. 
This criterion greatly reduces the computation load for discovering topological materials in a database of previously-synthesized compounds.  Of the more than 30,000 entires listed in a few selected nonsymmorphic space groups, our filling criterion alone eliminates 96\% of the entries before they are passed on for further analysis. From this guided search, we discover a handful of candidates, including the monoclinic crystals Ca$_2$Pt$_2$Ga, AgF$_2$, and Ca$_2$InOsO$_6$, and the orthorhombic crystal CsHg$_2$.
Based on \textit{ab initio} calculations, we show that these materials have filling-enforced Dirac nodes near the Fermi energy. In addition, we also identify CaPtGa as a promising filling-enforced Dirac-ring semimetal candidate.
\end{abstract}

\maketitle

\textit{Introduction.}---
Dirac and Weyl semimetals are three-dimensional analogues of graphene, where the conduction and valence bands touch at isolated points in the momentum space.\cite{PhysRevB.83.205101, PhysRevB.84.235126, Young2012}
The energy dispersion away from these points is linear in momentum, which leads to distinct electronic properties as compared to conventional  semiconductors with quadratically dispersing bands. They exemplify the large family of topological materials with nodal Fermi surfaces, which in 3D can compose of either points or lines with various geometries and dispersions.\cite{PhysRevB.83.205101, PhysRevB.84.235126, Young2012, PhysRevLett.112.036403, NComm_SrIrO3, Ring_Fu, PhysRevLett.116.186402,  Fang2016, bzduvsek2016nodal, Soluyanov2015, NewFermions, PseudospinVortex}
Nodal semimetals have intriguing experimentally observed physical properties, including ultrahigh mobility,\cite{Ong2015} protected surface states,\cite{Xu613} anomalous magneotoresistence potentially related to the chiral anomaly,\cite{Ong2015,PhysRevX.5.031023} novel quantum oscillations,\cite{Moll2016} and exotic optical properties.\cite{Orenstein2016}

Weyl semimetals require that either time-reversal (TR) $\mathcal T$ or inversion symmetry $\mathcal P$ be absent. In contrast, 3D Dirac semimetals may be symmetric under both $\mathcal T$ and $\mathcal P$, and a Dirac point is composed of a pair of Weyl nodes with opposite chirality protected by additional crystal symmetries. 
Such symmetry protection comes in two flavors. The first corresponds to accidental degeneracies along high-symmetry lines originating from band inversion.\cite{Yang2014} These accidental Dirac points can be moved along the symmetry direction and annihilate in pairs at time-reversal invariant momenta. 
The second type of Dirac semimetals features symmetry-enforced Dirac points that are naturally tuned near the Fermi energy by the electron filling.\cite{Young2012}
They represent the closer analogue to graphene, where the Dirac points are pinned to high-symmetry momenta, and therefore are generally well-separated in the Brillouin zone (BZ). 
Compared to the accidental Dirac semimetals created via band inversion, they are expected to be generically more isotropic and possess larger linearly-dispersing regions.

To date, there are two experimentally characterized Dirac semimetals: Na$_3$Bi and Cd$_3$As$_2$.\cite{Liu864,Neupane2014, Jeon2014, Liu2014} 
Both belong to the accidental class, and so far no experimental candidate has been found for the filling-enforced Dirac semimetals despite theoretical predictions.\cite{Young2012, PhysRevLett.112.036403, NewFermions}
More generally, while many theoretical possibilities are proposed and heroic efforts are made in their realizations, only a handful of materials systems, if any, are experimentally identified for each variant of nodal semimetals. 
Even these systems are often suboptimal in terms of physical properties, as ideally one would like the nodal points to have properties similar to those in graphene, which are well separated in the BZ and have significant velocities.

An important next step, crucial for moving nodal semimetals towards applications, is to identify general strategies for discovering new nodal semimetals.
A possible plan of attack is to perform data mining on databases of previously-synthesized 3D materials, in which one computes and examines the \textit{ab initio} band structure of every compound listed. 
Given the multitude of existing materials, general principles that can guide the quest are very valuable for such topological materials search.

Building upon a series of recent works on how nonsymmorphic spatial symmetries, which involve irremovable translations by fractions of the lattice constants, protect nodal semimetals,\cite{
PhysRevB.56.13607,
Young2012, PhysRevLett.112.036403, Sid2013, Ring_Fu, CavaDesign,usPNAS, Kee2016, Fang2016, PhysRevB.93.085427, PhysRevLett.116.186402,NComm_SrIrO3,SidLuttinger}
here we demonstrate how an electron-filling criterion, proven in Ref.~[\onlinecite{usPRL}] by some of us, can guide the search for nodal semimetals. The criterion uses only the symmetries and electron filling of the crystal, and applies to all 230 space groups (SGs) with or without spin-orbit coupling (SOC).
Only two properties of the system are required as input: $\nu_F$, the electron count in the (chemical) formula unit $F$, and $Z^*$, the multiplicity of $F$ in a symmetry-adapted `nonsymmorphic' unit cell.
The criterion states that if both $\nu_F$ and $Z^*$ are odd, a band insulator is forbidden.
In particular, when the SG is nonsymmorphic, the criterion forbids band insulators even when the Fermi surface volume vanishes. Nodal semimetals represent the simplest fermiology satisfying these conditions, i.e. they have zero Fermi volume but at the same time are {\em not} insulators, and we refer to such systems as filling-enforced semimetals (feSMs). 

Utilizing this general filling criterion, one can significantly narrow down the search space using only a minimal amount of information.
Here, we report the result of the first phase of our search within the Inorganic Crystal Structure Database (ICSD) of experimentally-characterized compounds,\cite{ICSD} where we focus on relatively simple SGs that can host Dirac feSMs.
Specifically, we consider the four monoclinic, centrosymmetric SGs {\bf \emph{11}}, {\bf \emph{13}}, {\bf \emph{14}} and {\bf \emph{15}}.
In addition, we also perform the search in a particular SG, {\bf \emph{74}} (\textit{Imma}), which is special as its symmetries are compatible with a minimal Dirac feSM with a Fermi surface comprising of only two Dirac points that are both filling-enforced and symmetry-related.\cite{Thesis}

Together with some additional filters to screen out strongly correlated materials, we significantly reduce the number of \emph{ab initio} density-functional theory (DFT) calculations required in the search.
From the search we discovered 13 candidates distributed among four different isostructural Dirac feSM materials families, which are respectively represented by {\protect \CGP}, AgF$_2$, Ca$_2$InOsO$_6$, and CsHg$_2$. 
As DFT calculations were only performed on 154 materials candidates, this gives a yield of $8\%$.\footnote{
Note that such yield is not a sharply defined concept, as real materials do not realize the nodal features in an ideal manner, and consequently a certain degree of arbitrariness is involved in judging whether or not a given system is a good feSM candidate.
} 

Among the Dirac feSM candidates we found, the air-stable compound {\protect \CGP}\cite{CGP-str} is particularly promising, as it does not have obvious tendency toward magnetic ordering. Based on DFT results, we find that  {\protect \CGP} hosts four distinct Dirac points near the Fermi energy, where two of them are filling-enforced, and the other two are accidental and form a symmetry-related pair. Their Dirac velocities are on the order of $10^5$ m/s, which are on par with those reported for Cd$_3$As$_2$ and Na$_3$Bi. In addition, one of the Dirac points is filling-enforced and well-isolated, and as expected is more isotropic than the accidental ones in Cd$_3$As$_2$ and Na$_3$Bi.
Finally, motivated by the discovery of the Dirac feSM {\protect \CGP}, we also study its structural and chemical cousins and found that CaPtGa, which crystalizes in SG {\bf \emph{62}} (\textit{Pnma}), is a Dirac-ring feSM candidate.

Before we move on to present our results, we discuss the relation between the present work and earlier studies invoking related ideas.
The possibility of filling-enforced Dirac semimetals was proposed in Refs.~[\onlinecite{Young2012}] and [\onlinecite{PhysRevLett.112.036403}], but there the discussion was motivated from the analysis of a particular model Hamiltonian, and only hypothetical candidates were proposed. Whether they can be experimentally synthesized is still an open challenge.
In contrast, Ref.~[\onlinecite{CavaDesign}] studied experimentally-characterized materials based on a design principle similar to our criterion in spirit, in that it also utilizes electron count and nonsymmorphic symmetries. However, the principle there does not properly account for the effect of formula-unit multiplicities, and hence can lead to `false positives'. For instance, the compound Cr$_2$B proposed in their work is \emph{not} filling-forbidden from being a band insulator.
A more recent work, Ref.~[\onlinecite{NewFermions}], briefly discussed identifying potential feSMs based on the criterion in Ref.~[\onlinecite{usPRL}], but there only one Cu-based candidate is proposed. That candidate is likely to be magnetically ordered at low temperature, and the applicability of the non-magnetic band-structure calculation is unclear. In contrast, magnetic ordering is considered unlikely in {\protect \CGP}  .

\textit{Materials search.}---
We will begin by discussing the filling criterion used to guide our materials search. Consider a stoichiometric compound with full crystalline order.
A minimal characterization of its structure involves three pieces of data: (i) the chemical formula unit $F$,  (ii) the number of times $F$ is repeated in the unit cell, commonly denoted by $Z$, and (iii) the space group $\mathcal G$ describing the symmetries of the crystal.
The electron count in $F$, $\nu_F$, can be readily found by summing over the atomic numbers involved. Although this count will include tightly-bound closed-shell electrons in $F$, this does not affect the conclusion of the filling criterion.

The next step in applying the criterion is to evaluate $Z^*$, which is defined as a refined notion of $Z$ with respect to a symmetry-adapted `nonsymmorphic' unit cell.
This can be efficiently achieved via the notion of `Wyckoff positions' $\mathcal W^{\mathcal G}_{w}$.
Given an SG $\mathcal G$, points in space are grouped into Wyckoff positions, indexed by $w=\rm{a},\rm{b},\dots$, which can be viewed as a tabulation of all the possible $\mathcal G$-symmetric lattices.\cite{ITC} Each Wyckoff position has a multiplicity $|\mathcal W^{\mathcal G}_{w}|$, defined as the minimum number of sites per unit cell needed for any lattice in that Wyckoff position. 
Wyckoff positions are labeled in a way such that points in $|\mathcal W^{\mathcal G}_{\rm a}|$ always have the highest site symmetry, and as a result $|\mathcal W^{\mathcal G}_{w}| \geq |\mathcal W^{\mathcal G}_{\rm a}|$ for any $\mathcal G$. In fact, except for four exceptional SGs ({\bf  \emph{199}}, {\bf  \emph{214}}, {\bf  \emph{220}} and {\bf  \emph{230}}), which we refer to as `Wyckoff-mismatched',\cite{feQBI} the multiplicities of the Wyckoff positions are always integer multiples of the smallest one, i.e. they satisfy $|\mathcal W^{\mathcal G}_{w}| = n_w |\mathcal W^{\mathcal G}_{\rm a}|$ for some positive integer $n_w$.

From definitions, $|\mathcal W^{\mathcal G}_{\text{a}}|>1$ if and only if $\mathcal G$ is nonsymmorphic, as the unit cell contains a point-group origin if and only if $\mathcal G$ is symmorphic.
This suggests that $|\mathcal W^{\mathcal G}_{\text{a}}|$ encodes the nonsymmorphic properties of $\mathcal G$, and indeed one can show that for most SGs the volume of the `nonsymmorphic unit cell', which enters the defination of $Z^*$, is $1/|\mathcal W^{\mathcal G}_{\text{a}}|$ of the primitive one.\cite{usPNAS, usPRL}
More precisely, the filling criterion can be tersely summarized as follows: A crystal, symmetric under time-reversal (TR) and a space group $\mathcal G\neq${\bf  \emph{199}}, {\bf  \emph{214}}, {\bf  \emph{220}} or {\bf  \emph{230}},\footnote{
Note that we exclude the four Wyckoff-mismatched space groups only for simplicity, and the filling criterion still holds with a slight modification in the definition of $Z^*$: instead of $|\mathcal W^{\mathcal G}_{\rm a}|$, the highest-common factor of $\{ |\mathcal W^{\mathcal G}_w|~:~w={\rm a},{\rm b},\dots\}$ enters into the definition.
}
is forbidden from being a band insulator if
\begin{equation}
\text{\bf both } \, \nu_F \, \, \text{and}\,  Z^*=Z/|\mathcal W^{\mathcal G}_{\text{a}}| \, \text{ are {\bf odd}.}
\label{condition}
\end{equation}

While the filling criterion alone filters out all band insulators from the search (alongside with metals or even nodal semimetals that are not filling-enforced), we apply additional filters before performing the band-structure calculations to improve the search efficiency. First, as the filling criterion is only applicable to weakly-correlated materials with TR symmetry, we remove materials in the list containing the magnetic atoms Mn, Fe, Co and Ni, or those containing elements with atomic numbers in the ranges 58-69 and 91-118, which contain valence f-shell electrons. In addition, to reduce the computation load we impose an \textit{ad hoc} cutoff and keep only materials with $\nu_F \leq 300$. Finally, we note that the filters discussed thus far only utilize the minimal set of data $(F,Z, \mathcal G)$, and as a consequence the list will generally contain materials that do not process perfect crystalline order (i.e.~the atoms are not perfectly ordered in space despite they enter $F$ in integral ratios). Therefore, in practice one has to further check the structural data and screen out such compounds before performing DFT calculations. 

\begin{figure}[h]
  \centering
\includegraphics[width=0.45\textwidth]{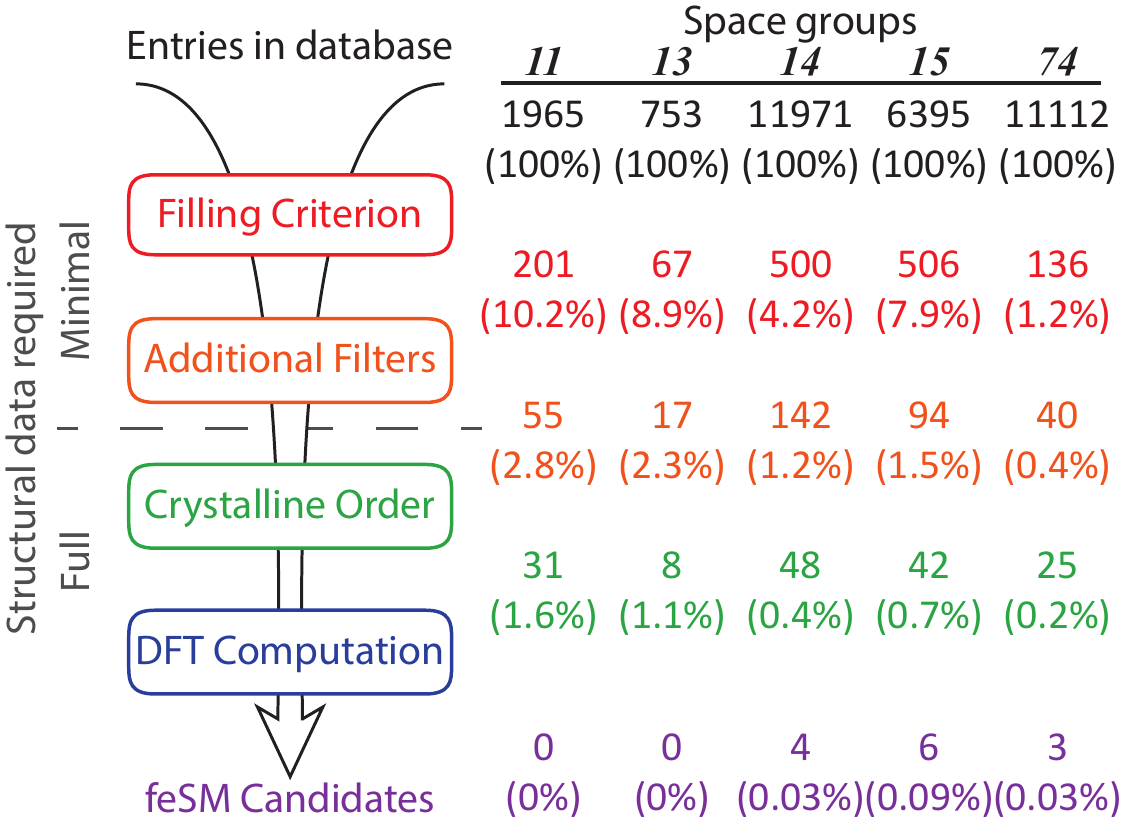}
\caption{
{\bf Guided search for filling-enforced semimetals using the electron filling criterion.} Entires in the materials databases, which correspond to previously-synthesized compounds, are passed through three filters to identify materials candidate for density functional theory (DFT) calculations. The first two stages use only a minimal set of structural data characterizing the materials, and therefore large scale screening is possible. 
We consider materials data belonging to five particular space groups in the ICSD.
The table shows the number of entries retained after each stage,
The first filter, corresponding to the filling condition (\ref{condition}), filters out all band insulators from the search and by itself reduces the materials search space to $\leq 10 \%$ of the original.
Note that the number of candidates (last row) will depend on the standard one imposes on the quality of the candidate, say the proximity of the nodes to the native chemical potential, and here we simply quote the number of candidates we discuss in this work.
}
\label{fig:Stat}
\end{figure}

The performance of the search protocol is summarized in Fig.~\ref{fig:Stat}. Among the five SGs we studied, the filling criterion alone reduces the number of candidates to $\leq 10\%$ of the entries listed on ICSD. The additional filters and crystalline order check together reduce the number to roughly $1\%$. DFT calculations are then performed on these candidates. While these systems typically exhibit nodal features near the Fermi energy, the majority of them are still plagued by a large density of states coming from multiple electron and hole pockets, or have negligible SOC and consequentially shows symmetry-enforced band sticking around the putative Dirac points. Nonetheless, from the search we identify four promising Dirac feSM candidates, each representing one  isostructural family of feSM materials, which leads to a total of 13 candidates discovered.
Note that this amounts to a yield of $8 \%$ relative to the number of DFT calculations performed.

\begin{figure}[h]
  \centering
\includegraphics[width=0.45\textwidth]{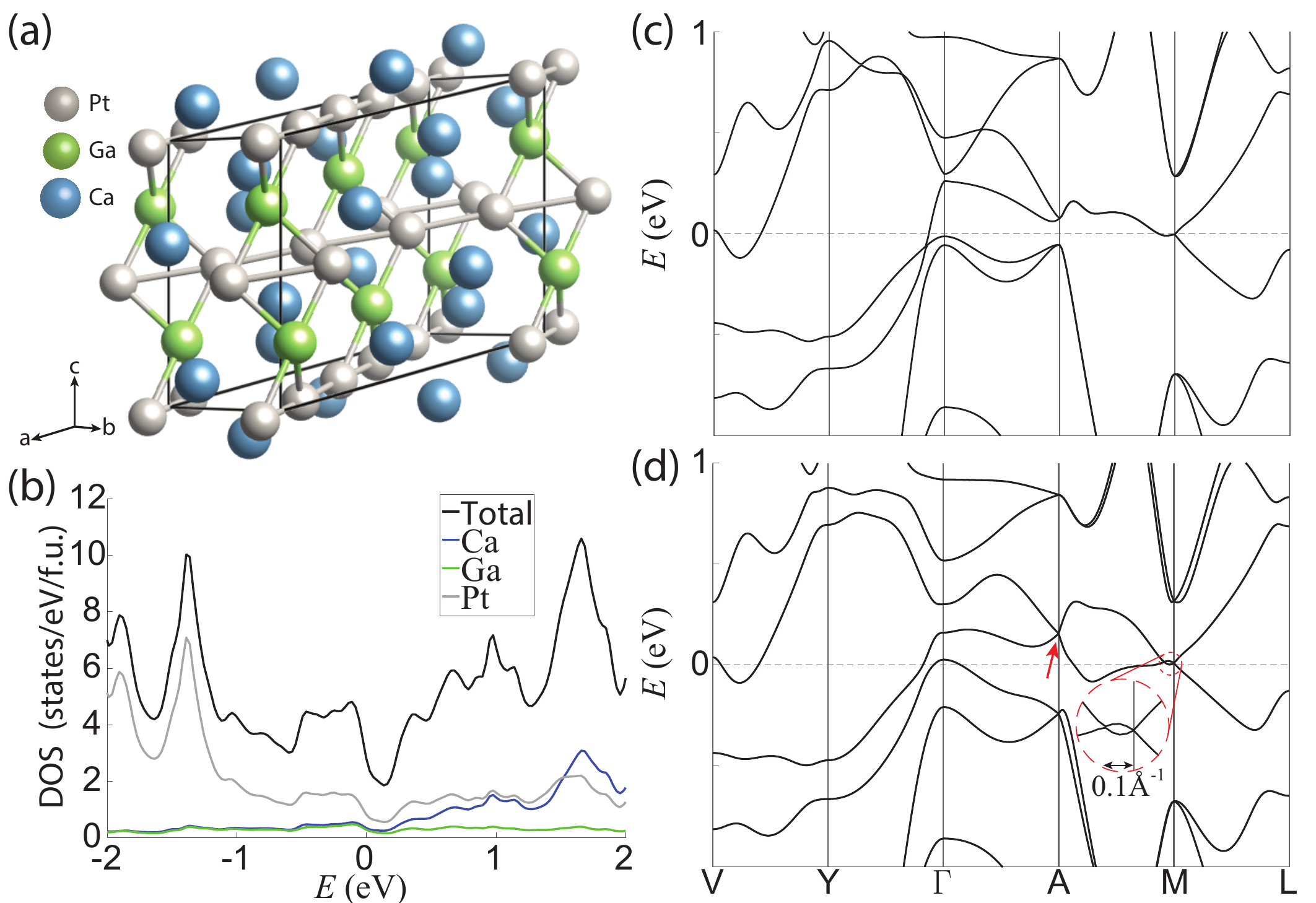}
\caption{
{\bf Filing-enforced Dirac semimetal candidate {\protect \CGP}.}
(a) Structure of {\protect \CGP}, which crystalizes in space group {\bf \emph{15}} ($C2/c$).\cite{CGP-str,Vesta}
(b) Projected density of states indicating semimetallic behavior.
(c,d) Band structures without (c) and with (d) spin-orbit coupling incorporated. Spin-orbit coupling lifts the line-degeneracy and leads to two filling-enforced Dirac points at A and M near the Fermi energy.
Inset shows a close-up for the band structure near M, where an additional accidental Dirac point is found.
}
  \label{fig:CGP}
\end{figure}

\textit{Dirac feSM candidate \CGP.}---
We begin by discussing the best Dirac feSM candidate we found: the intermetallic compound {\protect \CGP}, which was reported to crystalize in the Ca$_2$Ir$_2$Si-type structure in Ref.~[\onlinecite{CGP-str}] (Fig.~\ref{fig:CGP}a). 
The crystal is base-centered monoclinic with SG {\bf \emph{15}} ($C2/c$). In particular, SG {\bf \emph{15}} contains a glide and is hence nonsymmorphic, which is reflected in the minimum Wyckoff multiplicity $| \mathcal W_{\rm a}^\text{\bf \emph{15}}|=2$ (measured with respect to the primitive cell). 
The crystal structure of {\protect \CGP} can be understood as a 3D network formed by Ga and Pt, with interstitial space occupied by the Ca atoms.
The Pt atoms form chains running parallel to the $ab$ plane, with alternating layers of chains along the $c$ direction related by the glide. The Pt chains are then connected by Ga atoms, each four-coordinated with the Pt atoms, sitting midway between the layers.
Each primitive unit cell contains two formula units ($Z=2$), and therefore both $\nu_{\rm F}$ and $Z/| \mathcal W_{\rm a}^\text{\bf \emph{15}}|=1$ are odd, implying {\protect \CGP} is a feSM candidate.

To study whether or not {\protect \CGP} is a Dirac feSM, we perform non-spin-polarized electronic structure calculations using DFT within the generalized gradient approximation (PBE-GGA),\cite{gga} as implemented in VASP.\cite{vasp,vasp2} We use a plane wave cutoff of $280$ eV and a k-point grid for BZ integrations of $12 \times 12  \times 8$. Experimental structural data\cite{CGP-str} is used.
From the projected density of states, we find that the character of the {\protect \CGP} band structure near the Fermi energy has contributions from all three atomic species (Fig.~\ref{fig:CGP}b). 
At the Fermi energy, there is a sharp reduction of the total density of states, suggesting {\protect \CGP} is semimetallic. The band structure in the absence of SOC is shown in Fig.~\ref{fig:CGP}c. A continuous gap between the conduction and valence bands is found everywhere except for the A-M branch, which exhibits symmetry-enforced `band sticking'.

Thanks to the significant SOC in Pt, the undesirable line-degeneracy is lifted when SOC is incorporated (Fig.~\ref{fig:CGP}d). This gives rise to two filling-enforced Dirac points at momenta $A$ and $M$, which are at {$0.16$ eV} and {$0.01$ eV} above the Fermi energy respectively.
Generically, a Dirac point is characterized by a `tilt' vector $\vec t$ and three Dirac speeds $v_i$ characterizing the linear dispersion along three orthogonal directions.
However, symmetry constraints at the filling-enforced Dirac points A and M forces $\vec t_{\rm A, \rm M}=\vec 0$, and therefore these two Dirac points are characterized only by the Dirac speeds, which are respectively $(v_1,v_2,v_b) = (3.59, 1.19, 2.54) \times 10^5$ m/s for A and $(v_1,v_2,v_b) = (6.22, 0.86, 0.42) \times 10^5$ m/s for M. Here, $v_1$ and $v_2$ are along two orthogonal directions on the $k_a$-$k_c$ plane and $v_b$ is along the $k_b$ direction connecting A and M (Appendix \ref{app:DiracPt}).

Note that the Dirac point at M is very anisotropic, and this arises partly as a consequence of its proximity to another (TR-related) pair of accidental Dirac points originating from the crossing between the conduction and valence bands along the A-M branch (Fig.~\ref{fig:CGP}d). The crossing, protected by a two-fold rotation symmetry (Appendix \ref{app:DiracPt}), corresponds to an accidental Dirac point, which is located at $0.04\times \frac{4\pi}{b} = 0.1\,\AA^{-1}$ away from M and at $0.01$ eV above the Fermi energy. 
For this Dirac point, the tilt vector $\vec t$ is symmetry-pinned to lie along the $k_b$ direction, and its magnitude is found to be $0.21 \times 10^5$ m/s; the three Dirac speeds are $(v_1,v_2,v_b) = (7.74, 1.01, 0.33) \times 10^5$ m/s.
We also generated band structures using the HSE hybrid functional.\cite{hybrid} Our HSE calculations lead to the same conclusions, the same Dirac points, and similar associated energy scales comparing to those obtained with GGA-PBE (Appendix \ref{app:HSE}).

These observations suggest that {\protect \CGP} is an interesting Dirac system featuring both filling-enforced and accidental Dirac points near the Fermi energy, where the magnitude of their Dirac velocities are all on par with the current experimental candidates.
In particular, the Dirac point at A is very isotropic, as one would expect for a filling-enforced and well-isolated Dirac point.
This can be seen from the anisotropy parameter $\alpha \equiv \text{max}\{v_i\}/ \text{min}\{v_i\}$, where $i$ runs over the three independent linear speeds.
The values of $\alpha$ are $3.0$, $14.8$ and $23.5$ for the Dirac points at A, M and on A-M respectively, which can be contrasted with the experimentally-measured values for Na$_3$Bi ($\alpha = 4.4$)\cite{Liu864} and Cd$_3$As$_2$ ($\alpha = 4.0$)\cite{Liu2014}.

\textit{Other Dirac feSM candidates.}---
Next, we present our results on the other Dirac feSM candidates discovered. As discussed, the 13 materials discovered fall into four isostructural families, which we tabulate in Table \ref{tab:FullList} (Appendix \ref{app:Families}). {\protect \CGP} represents one of the families, and here we discuss the representatives of the other three. 
Compared to {\protect \CGP}, however, these materials either show higher tendency to magnetic ordering, or have smaller Dirac speeds due to a smaller SOC lifting of band sticking. 

\begin{center}
\begin{table}[h]
\caption{Table of isostructural families of Dirac feSM candidates discovered in the search.\label{tab:FullList}}
\begin{tabular}{ccc}
SG ~&~ Representative ~&~ Others\\
\hline
14 ~&~ AgF$_2$ ~&~ CuF$_2$\\
14 ~&~ Ca$_2$InOsO$_6$  ~&~ Sr$_2$InOsO$_6$\\
15 ~&~Ca$_2$Pt$_2$Ga~&~Ca$_2$Pd$_2$Ga, Ca$_2$Pt$_2$In, Ca$_2$Pd$_2$In,\\
~&~&Sr$_2$Pt$_2$In, Sr$_2$Pd$_2$In\\
74 ~&~ CsHg$_2$ ~&~ KHg$_2$, RbHg$_2$
\end{tabular}
\end{table}
\end{center}

\begin{figure}[bth]
  \centering
\includegraphics[width=0.5\textwidth]{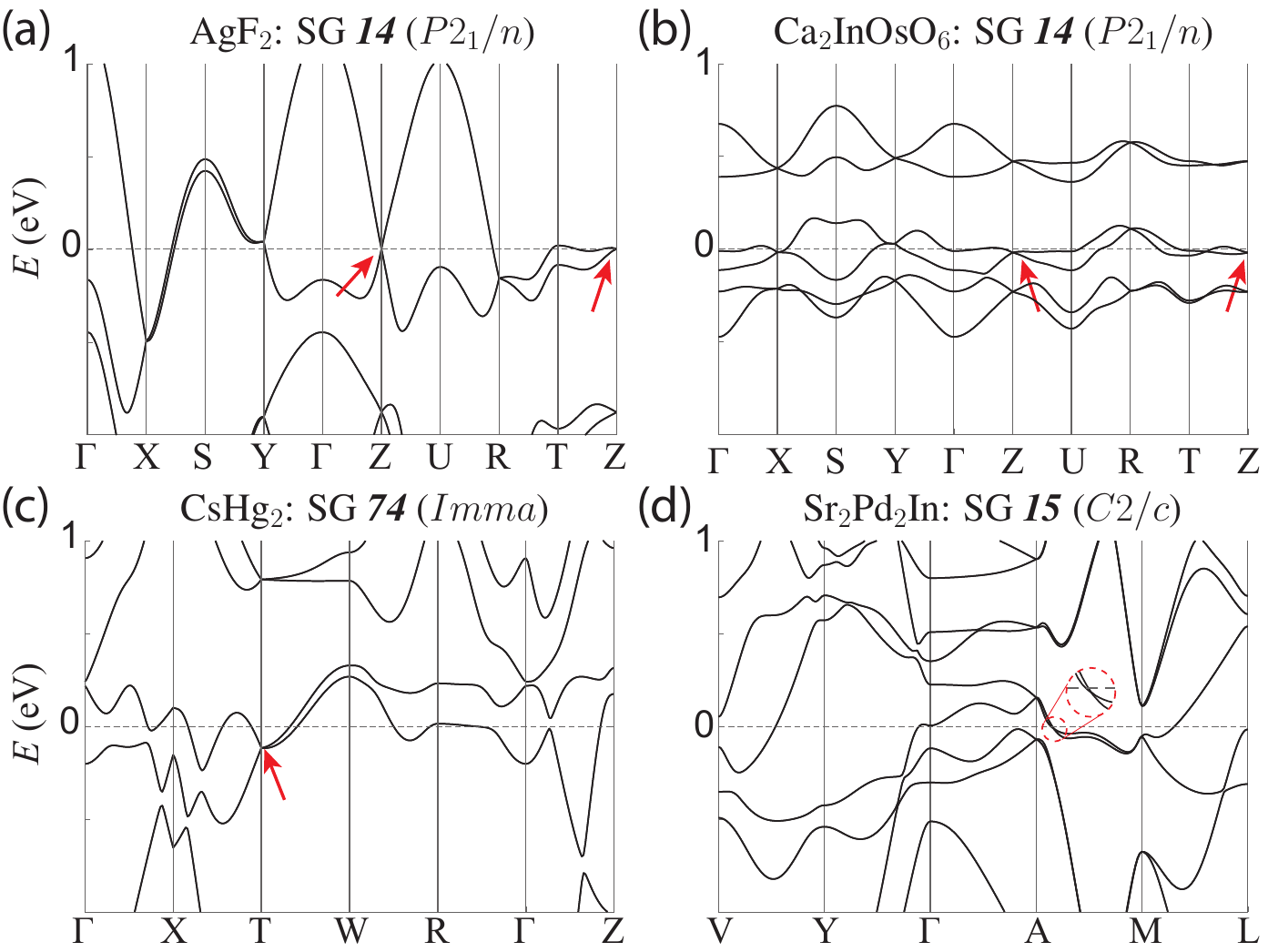}
\caption{
{\bf Band structures for additional filling-enforced Dirac semimetal candidates.}
The dominant filling-enforced Dirac points are indicated by red arrows. 
Note that these are results from non-spin-polarized band structure calculations.
However, the presence of d-shell atoms in (a-b) may lead to magnetically ordered ground states, in which case the band structures will need to be recalculated and the filling constraints may also be modified depending on the nature of symmetry lowering associated with the magnetism. (c) For space group {\bf \emph{74}}, the filling constraints can be satisfied by a minimal Dirac semimetal with a pair of symmetry-related Dirac points at T and its symmetry partner (not shown).
(d) Although the filling-enforced Dirac points at A and M have small Dirac speeds along the A-M branch, an accidental type-II Dirac point is found to sit close to the Fermi energy.
}
\label{fig:Others}
\end{figure}

Similar to before, non-spin-polarized calculations were performed using experimental structural data from ICSD as input.
First, we discuss the two families found in the centrosymmetric monoclinic SG {\bf \textit{14}} ($P2_1/n$), represented by AgF$_2$ and Ca$_2$InOsO$_6$. 
As one can see from Figs.~\ref{fig:Others}(a,b), energy bands symmetric under SG {\bf \textit{14}} are four-fold degenerate at four time-reversal invariant momenta. For the two candidates at hand, the filling-enforced Dirac point at Z sits close to the Fermi energy and the valence-band Dirac velocities are significant.
While the electronic band structures look promising, we caution that these candidates contain valence d-shell electrons, and are therefore susceptible to magnetic instability.

We also perform the guided search within materials with SG {\bf \textit{74}} ($Imma$), which is one of the SGs capable of hosting a minimal Dirac feSM with only two symmetry-related filling-enforced Dirac points.\cite{Thesis} Among them, we identify CsHg$_2$ as the most promising candidate, where the filling-enforced Dirac point at T is at $0.12$ eV below the Fermi energy (Fig.~\ref{fig:Others}c).  However, the Dirac speed along the T-W branch almost vanishes, as that branch exhibits band sticking in the absence of SOC.\cite{BnC}

Finally, in Fig.~\ref{fig:Others}d we plot the band structure for the compound Sr$_2$Pd$_2$In, which belongs to the {\protect \CGP} family.
Although the filling-enforced Dirac points at A and M suffer from a small SOC-lifting of band degeneracy, we found that an accidental type-II Dirac point is located near the Fermi energy, and hence this compound serves as a promising type-II Dirac semimetal candidate.

\begin{figure}[tbh]
  \centering
\includegraphics[width=0.45\textwidth]{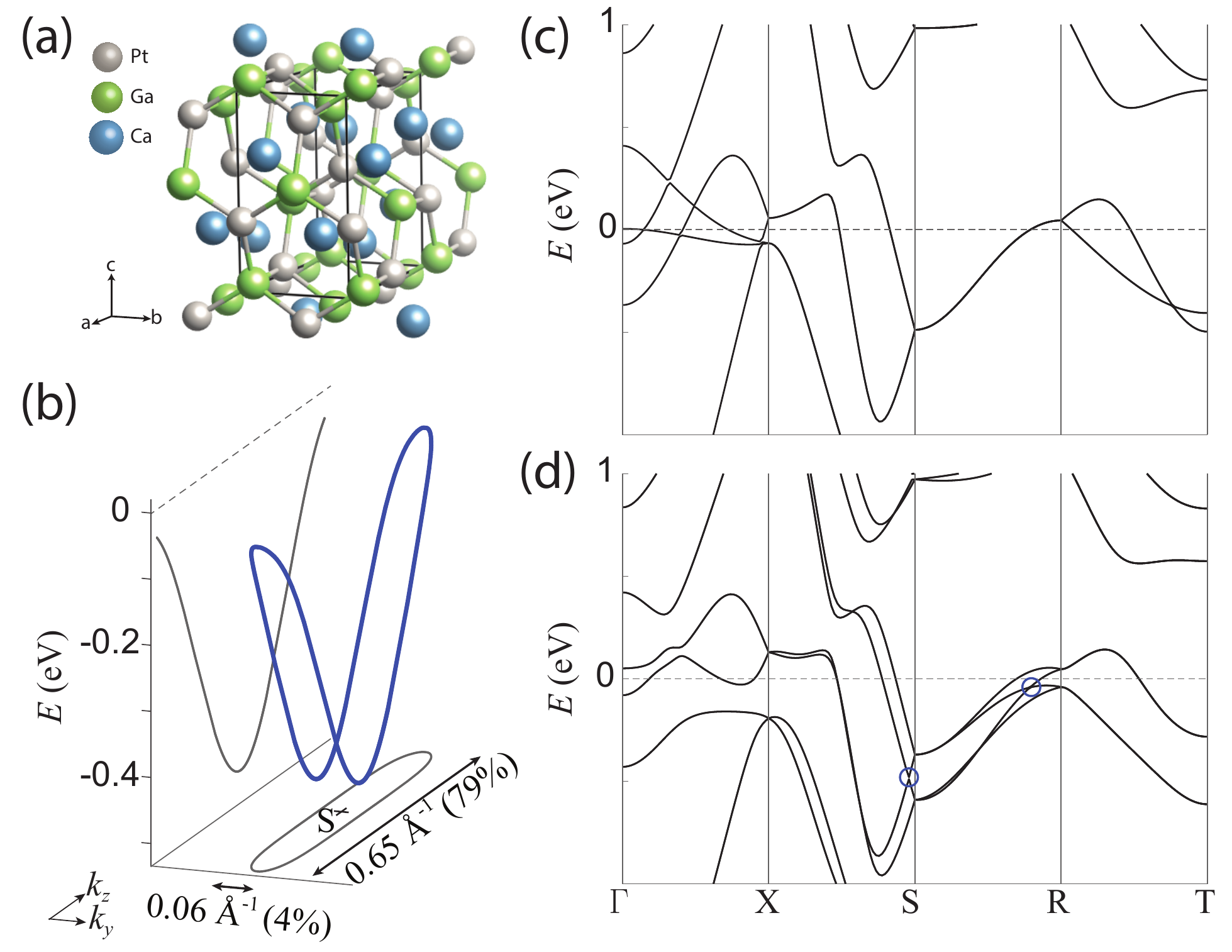}
\caption{
{\bf Characterization of the filing-enforced Dirac-ring semimetal candidate CaPtGa.}
(a) Crystal structure of CaPtGa in space group {\bf \emph{62}} ($Pnma$).\cite{CGP-line_str,Vesta}
(b) Energy-momentum plot of the Dirac ring (blue). The ring lies on the $k_y$-$k_z$ plane and encircles point S. Gray curves show projection of the ring along the corresponding direction.
(c,d) Band structure without (c) and with (d) spin-orbit coupling incorporated. Spin-orbit couping lifts band degeneracies along X-S and S-R, leaving behind four-fold band crossing corresponding to the intersection of the high-symmetry lines and the Dirac ring (blue circles).
}
  \label{fig:CGP-line}
\end{figure}

\textit{Dirac-ring feSM candidate CaPtGa.}---
Having identified certain materials systems as being promising feSM candidates, it is natural to explore if their structurally- and/ or chemically-related systems are also feSM candidates. In this process, we also discovered a Dirac-ring feSM CaPtGa, which is chemically similar to the Dirac feSM {\protect \CGP} but crystallizes in a centrosymmetric orthorhombic structure with SG {\bf \emph{62}} (\textit{Pnma}).\cite{CGP-line_str} The crystal structure of CaPtGa can again be understood as a 3D network of Ga and Pt atoms with Ca atoms occupying the interstitial space (Fig.~\ref{fig:CGP-line}a). The GaPt network can be viewed as a distorted layered structure stacked along the $a$ direction, where in each layer the Ga and Pt atoms form honeycomb lattices akin to that of hexagonal boron nitride, and the interlayer Ga-Pt bond lengths are strongly modulated to render the structure three-dimensional.

From a symmetry and filling perspective, CaPtGa is identical to SrIrO$_3$, which crystalizes in the same SG. As SrIrO$_3$ is proposed to be a topological Dirac-ring feSM candidate,\cite{NComm_SrIrO3,Ring_Fu, usPRL} one would naturally expect the same for CaPtGa (Appendix \ref{app:Ring}). This is confirmed by our DFT calculations, where a Dirac-ring is found on the plane $(\pi, k_y, k_z)$ encircling point S$\equiv(\pi,\pi,0)$. 
\footnote{Note that Ref.~[\onlinecite{NComm_SrIrO3}] reports the crystal structure in a different setting (\textit{Pbnm}), which leads to a different parameterization of the BZ.}
As shown in Fig.~\ref{fig:CGP-line}b, the nodal ring is elongated along the $k_z$ direction, with an aspect ratio of $1:10.8$. However, it has a significant energy dispersion and spans an energy range of $0.44$ eV.
Note that SOC is essential to the existence of the nodal ring, as the bands along S-R are eight-fold degenerate in the absence of SOC (Fig.~\ref{fig:CGP-line}c and d).

\textit{Discussion}---
Using the filling criterion developed in Ref.~[\onlinecite{usPRL}], we perform a directed search for nodal semimetals and identify {\protect \CGP}, AgF$_2$, Ca$_2$InOsO$_6$, and CsHg$_2$, representing four different isostructural families, as good Dirac feSM candidates. {\protect \CGP} is found to be particularly promising, and features both filing-enforced and accidental Dirac points near the Fermi energy. We also identify the chemically-similar compound CaPtGa as a Dirac-ring feSM candidate. 

The present work suggests multiple directions for future research. On the experimental side, it is of immense interest to synthesize the reported materials and study their physical properties; on the theoretical side, the search protocol can also be applied to identify new Weyl semimetal candidates in inversion-breaking SGs, or more generally one can hunt for all flavors of feSMs by an exhaustive search in all SGs. 
It can also be used as a guide for the design of new materials with desired properties.
Finally, the filling criterion can be generalized to magnetic space groups and aid the search for feSMs with broken TR symmetry. \\~\\

\begin{acknowledgments}
We thank Haruki Watanabe and Michael P. Zaletel for collaboration on earlier works and useful comments on the manuscript.
We also thank Tess Smidt for helpful discussions.
RC and JBN were supported by the Laboratory Directed Research and
Development Program of Lawrence Berkeley
National Laboratory under DOE Contract No. DE-AC02-05CH11231.
Work at the Molecular Foundry was supported by the Office of Science,
Office of Basic Energy Sciences, of the U.S. Department of Energy, and
Laboratory Directed Research and Development Program at the Lawrence
Berkeley National Laboratory, under Contract No. DE-AC02-05CH11231.
AV and HCP were supported by NSF DMR-141134 and ARO MURI program W911NF-12-1-0461.
We also thank NERSC for computational resources. AV and RC were also partly  funded by the U.S. Department of Energy, Office of Science, Office of Basic Energy Sciences, Materials Sciences and Engineering Division under Contract No. DE-AC02-05-CH11231 (Quantum Materials program KC2202).
\end{acknowledgments}

\bibliography{references}
\clearpage

\onecolumngrid

\begin{appendix}

\section{Symmetry analysis of the Dirac points
\label{app:DiracPt}}
Generically, the band dispersion about a Dirac point can be expanded to leading order as $\delta E_{\delta \vec k}^{\pm} = \vec t \cdot \delta \vec k \pm \sqrt{\sum_{i,j=1}^3 A_{ij}\delta k_i \delta k_j}$, where $A_{ij}$ is a real symmetric matrix. The square root of the three eigenvalues of $A_{ij}$ are the Dirac speeds characterizing the shape of the Dirac cone, and are generally independent of each other; the vector $\vec t$ parameterizes the `tilt' of the cone, which is generically allowed as condensed matter systems do not possess genuine Lorentz invariance.\cite{Soluyanov2015} Here, we discuss the symmetry constraints on the Dirac-point parameters for {\protect \CGP}. The corresponding analysis for the other Dirac feSM materials will be similar.

While the existence of Dirac points at A and M can be inferred from earlier works,\cite{BnC, Thesis} we perform a detailed $\vec k \cdot \vec p$ analysis here to demonstrate how these Dirac points emerge.
SG {\bf \emph{15}} is generated by the lattice translations, inversion $\mathcal P$ about the origin, and a $\pi$-rotation about the axis $(0,y,1/4)$, $\mathcal R_{y} =\{ R_y ~|~ (0,0,1/2)\} $. 
Note that the combined symmetry $\mathcal G_y\equiv \mathcal R_y \mathcal P = \{ m_y ~|~ (0,0,1/2) \}$ is a glide with $x$-$z$ being the mirror plane.
The nonsymmorphic nature of the SG leads to additional band degeneracy at the TR invariant momenta A $\equiv (0,0,\pi)$ and M $\equiv(0,2\pi,\pi)$,\cite{BnC,Thesis} which are both symmetric under all spatial symmetries. (We parameterize the BZ using the coordinate system reciprocal to the conventional unit cell, with the lattice constants suppressed, such that the reciprocal lattice vectors are $\vec G_a = 2\pi (1,1,0)$, $\vec G_b = 2\pi (1,-1,0)$ and $\vec G_c = 2\pi (0,0,1)$.)
To see this, observe that $\mathcal R_y \mathcal P \mathcal R_y^{-1} \mathcal P ^{-1} =\{ 1 ~|~(0,0,1)\}$, which gives the phase factor $e^{- i k_z}$ when acting on Bloch states. As $e^{- i k_z} = -1$ at A and M, the two symmetries anti-commute and lead to a two-dimensional irreducible representation. In addition, this 2D representation cannot be symmetric under TR: the doublet contains bands with opposite $\mathcal P$ eigenvalues, but the two bands forming a Kramers pair necessarily have equal $\mathcal P$ eigenvalues as $[\mathcal P,\mathcal T]=0$.
This leads to a four-dimensional co-representation,\cite{BnC,Thesis} which forbids {\protect \CGP} from being a band insulator, as it has a (total) filling of $Z \nu_F = 454 = 2 \mod 4$ electrons per primitive cell (including all tightly bound electrons). Note that this conclusion can also be viewed as a special instance of the more general filling constraint developed in Ref.~[\onlinecite{usPRL}].

While having a four-fold band degeneracy is a prerequisite, to claim discovery of a Dirac point one has to further establish a linear dispersion around the nodal point.
Although this can be accomplished via purely group-theoretic methods,\cite{Thesis,Young2012} we provide here an explicit argument via the construction of a $\vec k\cdot \vec p$ effective Hamiltonian. Let $\tau^\mu$, $\sigma^\mu$ be two sets of Pauli matrices with $\mu=0$ denoting the identity. The stated symmetry representation at A and M can be realized by the four-dimensional unitary matrices
\begin{equation}\begin{split}\label{eq:URep}
\begin{array}{ll}
U(\mathcal R_y) =  -i \sigma^0 \tau^2;& 
U(\mathcal P)=  \sigma^0  \tau^3;\\
U(\mathcal G_y) = \sigma^0  \tau^1;&
U(\mathcal T) =  -i \, \sigma^2 \tau^0,
\end{array}
\end{split}\end{equation}
where the anti-unitary TR symmetry is represented by $U(\mathcal T)$ followed by complex conjugation. (For notational simplicity, we write  $\sigma^\mu  \tau^\nu \equiv \sigma^\mu \otimes \tau^\nu$.)

The Bloch Hamiltonian about $\vec k_0 = \rm{A}$ or ${\rm M}$ can be expanded as $H_{\vec k_0 + \delta \vec k} = E_{\vec k_0} + h_{\mu,\nu}^i \sigma^\mu \tau^\nu \delta k_i + \mathcal O(\delta \vec k^2)$, where $h^i_{\mu,\nu}$ are real coefficients which are generically non-zero.
Symmetries, however, place constraints on their values, and one sees that the only non-zero coefficients are $h^{x,z}_{j,1}$ for $j=1,2,3$, and $h^y_{0,2}$. 
This gives the dispersion
\begin{equation}\begin{split}\label{eq:kdotp}
\delta E_{\vec k_0+\delta \vec k} \approx
\pm \sqrt{|  \vec h^x \delta  k_x+ \vec h^z \delta k_z |^2 + (h^y_{0,2} \delta k_y)^2},
\end{split}\end{equation}
where $\delta \vec k \equiv (\delta k_x, \delta k_y, \delta k_z)$ parameterizes deviation from the Dirac point $\vec k_0$, $\delta E_{\vec k_0+\delta \vec k} \equiv E_{\vec k_0+\delta \vec k}  - E_{\vec k_0} $, $h^{i}_{\mu,\nu}$ are real coefficients, $\vec h^{x,z} \equiv (h^{x,z}_{1,1},h^{x,z}_{2,1},h^{x,z}_{3,1})$, and each band is doubly degenerate. 
This implies the only possible soft direction (i.e.~the dispersion scales at least as $\mathcal O(\delta \vec k^2)$) is along $\delta k_y = 0$ and $\vec h^x \delta  k_x+ \vec h^z \delta k_z=0$. 
As the vectors $\vec h^x$ and $\vec h^z$ are generically \emph{not} parallel, no soft direction is expected and this proves that A and M are indeed Dirac points protected by crystalline symmetries.

Now if one moves away from the high-symmetry point along the line A-M, the only remaining spatial symmetry is the rotation $\mathcal R_y$. To determine if two sets of bands can cross, one has to check whether they carry the same or different symmetry eigenvalues. 
This can be read off from Eq.~\eqref{eq:URep}: the two bands related by $\mathcal P \mathcal T$ corresponds to the $\sigma$ degrees of freedom, which have the same eigenvalues under $U(\mathcal R_y)$. Naively, one may think that this is inconsistent with earlier classification results,\cite{Yang2014} which states that a two-fold rotation alone cannot protect such accidental crossing. To clarify such apparent discrepancy, we study the problem in more details below.

Consider Bloch states satisfying $\mathcal R_y |\psi_{\pm } \rangle = \pm i | \psi_{\pm } \rangle$, and check that
\begin{equation}\begin{split}\label{eq:}
\mathcal R_y \left( \mathcal P \mathcal T | \psi_{\pm} \rangle\right) =& 
- \mathcal P  \mathcal T \mathcal R_y | \psi_{\pm} \rangle\\
 =& - \mathcal P  \mathcal T (\pm i)| \psi_{\pm} \rangle =  \pm i \left( \mathcal P \mathcal T | \psi_{\pm} \rangle\right),
\end{split}\end{equation}
and hence the doubly-degenerate $\mathcal P\mathcal T$-paired bands carry the same rotation eigenvalue, as we claimed. This implies different sets of bands can carry different $\mathcal R_y$ eigenvalues, and when they do their crossings are symmetry-protected. 
Note that had $[\mathcal P, \mathcal R_y]=0$, the conclusion would be different as the bands could only anti-cross.\cite{Yang2014} Since the modification of the commutator comes from the nontrivial relative position of the inversion center and the rotation axis, even this accidental crossing owes its existence to the nonsymmorphic nature of the SG.\\

Finally, the $\vec k\cdot \vec p$ effective Hamiltonian about an accidental Dirac point $\vec k_0'$ along the line A-M can be readily found as before. The only non-zero coefficients are now $h^{x,z}_{j,1}$ for $j=1,2,3$, $h^{x,z}_{0,3}$, $h^y_{0,0}$, and $h^y_{0,2}$. This gives the band dispersion
\begin{equation}\begin{split}\label{eq:kdotp}
\delta E_{\vec k_0'+\delta \vec k} \approx &h^y_{0,0} \delta k_y 
\pm \sqrt{|  \vec h'^x \delta  k_x+ \vec h'^z \delta k_z |^2 + (h^y_{0,2} \delta k_y)^2},
\end{split}\end{equation}
where $\vec h'^{x,z} \equiv (h^{x,z}_{0,3},h^{x,z}_{1,1},h^{x,z}_{2,1},h^{x,z}_{3,1})$ is now a four-component vector. This is qualitatively the same as those about A and M, except for the new symmetry-allowed `tilting term' $h^y_{0,0}$ giving rise to the possibility of realizing a so-called type-II Dirac point,\cite{Soluyanov2015} as is found in the compound Sr$_2$Pd$_2$In discussed in the main text. 
In addition, we remark that the crossing of the energy bands along the A-M branch can be confirmed by studying the evolution of $R_y$ eigenvalues, and we have performed this check for Ca$_2$Pt$_2$Ga and SrPd$_2$In.

\section{Symmetry analysis of the Dirac-nodal ring
\label{app:Ring}}
The existence of the nodal ring can be understood as follows: SG {\bf \emph{62}} contains an n-glide $\mathcal G_n\equiv\{ m_x ~|~(1/2,1/2,1/2) \}$, which remains as a symmetry on the plane $(k_x=\pi, k_y, k_z)$. An energy eigenstate $| \psi_{\vec k,\pm }\rangle$ on the plane can therefore be labeled by the eigenvalue $\mathcal G_n | \psi_{\vec k,\pm }\rangle = \pm i e^{- i (k_y+k_z)/2} | \psi_{\vec k,\pm }\rangle $.  In addition, as $\mathcal G_n \mathcal P \mathcal G_n^{-1} \mathcal P^{-1} = \{ 1 ~|~ (1,1,1)\}$, one finds that the two degenerate states $|\psi_{\vec k,\pm }\rangle$ and $\mathcal P \mathcal T | \psi_{\vec k,\pm} \rangle$ have the same $\mathcal G_n$ eigenvalues, and therefore two sets of band doublets carrying different $\mathcal G_n$ eigenvalues will cross when they come close in energy, which is observed in Fig.~\ref{fig:CGP-line}d along the line S-R. In fact, such band crossing is mandated due to a pair-switching as the momentum radiates outward from the four-fold degenerate point S to the momentum lines U-R and U-X, both of which also demand four-dimensional co-representations along the entire line.\cite{BnC,NComm_SrIrO3,Ring_Fu} This leads to a symmetry-protected Dirac nodal ring encircling S.

\section{Robustness of the Dirac features in {\protect \CGP}
\label{app:HSE}}
\begin{figure}[h!]
\begin{center}
{\includegraphics[width=0.4 \textwidth]{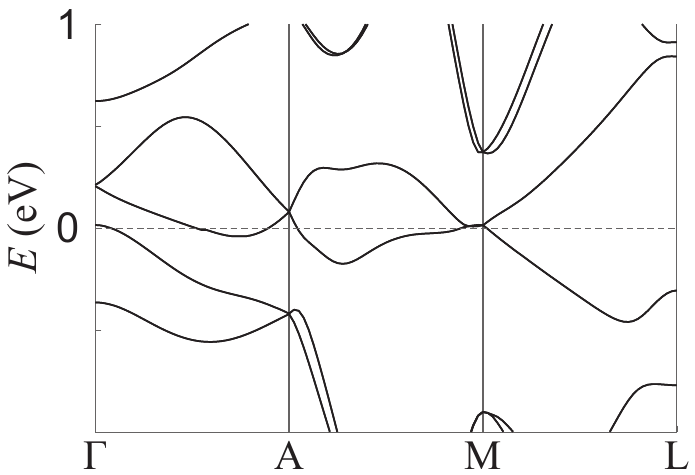}} 
\caption{Band structure of {\protect \CGP} computed using the HSE hybrid functional method.
\label{fig:HSE} 
 }
\end{center}
\end{figure}

To test whether all of the key feature of the Dirac points survive regardless of the methods of the \textit{ab initio} calculation, we also computed its band structure using the HSE hybrid functional method. We used $0.2$ for the range-separation parameter in range separated hybrid functionals and a $k$-point grid of $5 \times 5  \times  3$ for BZ integrations, due to the expensive cost of the hybrid functional HSE calculation. The band structure is shown in Fig.~\ref{fig:HSE}, from which we observed the same two filling enforced Dirac points and the accidental Dirac points. We notice that the filling enforced Dirac point at M and the accidental Dirac points are more anisotropic in this calculation.

\section{Additional Dirac feSM candidates
\label{app:Families}}
\begin{figure*}[h]
\begin{center}
{\includegraphics[width=1 \textwidth]{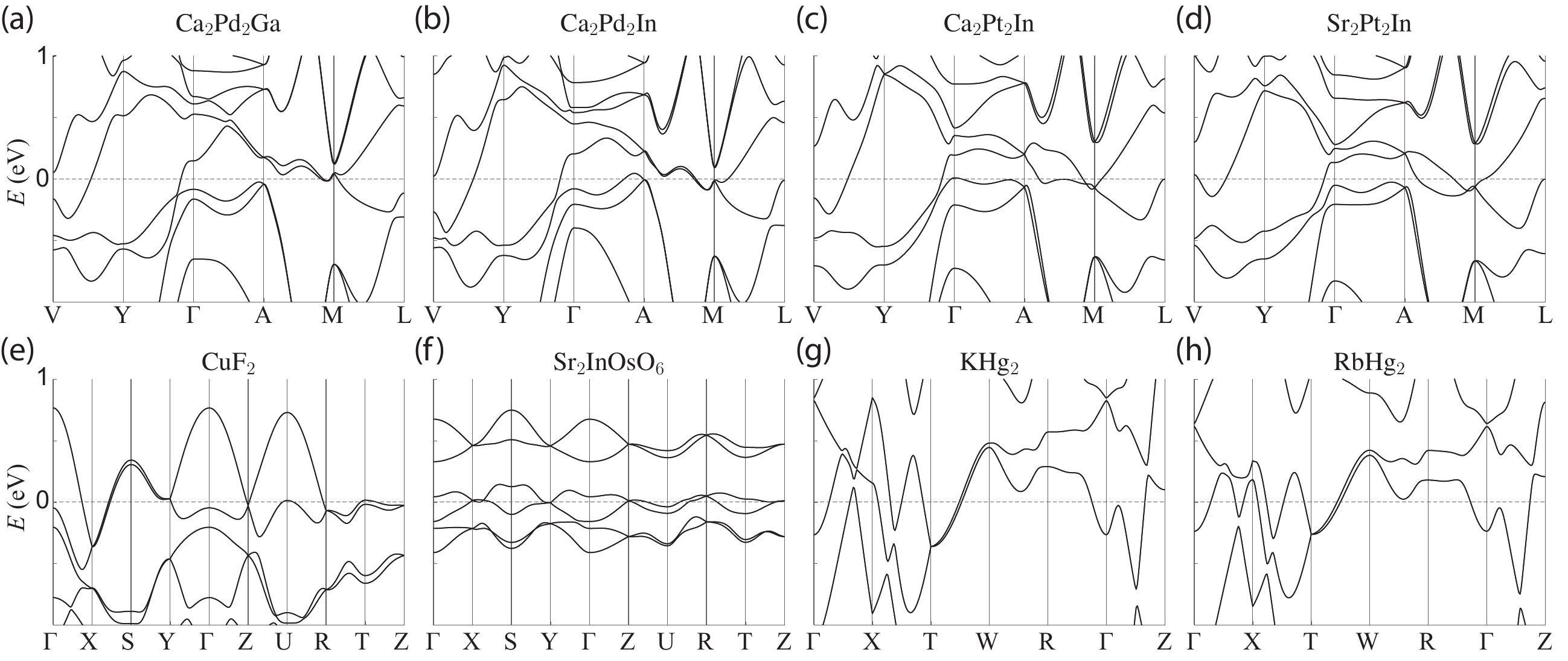}} 
\caption{Additional band structures for materials in the four isostructural families of Dirac feSMs. (a-d) Materials in the {\protect \CGP} family in SG {\bf \emph{15}} ($C2/c$). (e, f) Materials in the AgF$_2$ and Ca$_2$InOsO$_6$ families respectively, which are both in SG {\bf \emph{14}} ($P2_1/n$). (g-h) Materials in the CsHg$_2$ family in SG {\bf \emph{74}} ($Imma$).
\label{fig:supp_BS}
 }
\end{center}
\end{figure*}

In Table \ref{tab:FullList} we tabulated the 13 Dirac feSM candidates found in the four isostructural families. 
The band structures for five of them are discussed in the main text; in Fig.~\ref{fig:supp_BS} we present the results on the remaining ones.

\twocolumngrid
\end{appendix}

\end{document}